\documentclass[a4paper,11pt]{article}
\pdfoutput=1 \usepackage{jheppub} \usepackage[T1]{fontenc} 
\let\veps=\varepsilon
\newcommand{\sgn}{{\, \rm sgn}}

\renewcommand{\varepsilon}{{\, \rm sgn}}
\renewcommand{\Im}{{\rm Im}}

\def\beq{\begin{equation}}
\def\endq{\end{equation}}
\def\f{{h}}

\def\bR{{\mathbb R}}

\def\bZ{{\mathbb Z}}
\def\Im{\mathop{\rm Im}\nolimits}

\def\ff{{n}}

\title{\textcolor{black}{The massless thermal field and the thermal fermion bosonization in two dimensions\footnote{Giovanni Morchio (1948-2021). In memoriam.}}}

\author[a,b]{E. Akhmedov}
\author[c]{H. Epstein}
\author[c,d,e]{U. Moschella}

\affiliation[a]{Moscow Institute of Physics and Technology, Institutskii per. 9, 141700 Dolgoprudny, Russia}
\affiliation[b]{Institute for Theoretical and Experimental Physics, B. Cheremushkinskaya 25, 117218 Moscow, Russia}
\affiliation[c]{Institut des Hautes Etudes Scientifiques, 35 Route de Chartres, 91440 Bures-sur-Yvette}
\affiliation[d]{Universit\`a degli Studi dell'Insubria - Dipartimento DiSAT, Via Valleggio 11 - 22100 Como, Italy}
\affiliation[e]{INFN, Sez di Milano, Via Celoria 16, 20146, Milano, Italy}

\abstract{We come back to the issue of bosonization of fermions in two spacetime dimension and give a new costruction in the steady state case where left and right moving particles can coexist at  two different temperatures. A crucial role in our construction is played by translation invariant infrared states and the corresponding field operators which are naturally linked to the infrared behaviour of the correlation functions.  We present two applications:  a simple new derivation in the free relativistic case of a formula by Bernard and Doyon  and  a full operator solution of the massless Thirring model in the steady state case where  the left and right movers have two distinct temperatures.}

\makeatletter
\gdef\@fpheader{}
\makeatother

\begin{document}

\numberwithin{equation}{section}

\maketitle

\section{Introduction}

The theoretical study of  models of fermions in 1+1 space-time dimension has  regained interest and importance, mainly because  structures which are effectively one-dimensional are nowadays available and play a significant role in the current  technological development.

An important notion for  their understanding, at  both  relativistic and non relativistic level, is the so called  {\em bosonization}.  
This is  indeed is an extremely useful mathematical tool and, more importantly, 
it is a {\em fact} deeply rooted in the physical peculiarities of   one-dimensional fermionic systems.

In introducing his model \cite{tomonaga} Tomonaga   gives  credit to Bloch \cite{bloch1,bloch2} for the first observation of   {\em the fact  that in some approximate sense the behavior of an assembly of Fermi particles can be described by a quantized field of sound waves in the Fermi gas, where the sound field obeys Bose statistics}.  
Tomonaga's construction  later evolved in the Luttinger model, a model  of interacting fermions  linearized around the Fermi momenta \cite{luttinger}. 
The  zero temperature solution of the Luttinger model was given shortly after  by Lieb and Mattis  \cite{lieb}; the bosonization method allowed them to replace the fermionic Hamiltonian  by a quadratic expression in the boson collective modes and the charge operators and to diagonalize it by a Bogoliubov transformation. Actually, the description in terms of bosons revealed itself to be  appropriate to describe  the low energy excitations of a generic  interacting electron gas in one dimension - also called  a  "Luttinger liquid"  \cite{haldane}.  

The thermal equilibrium  correlation functions of  the Luttinger model were first constructed in \cite{hei} where again the  fermion bosonization procedure was a crucial ingredient. 
Since then, a considerable amount of work has been done to characterize the thermodynamical properties of one dimensional Luttinger liquids. In particular, intense study has been devoted   to   systems out of equilibrium whose ends are in contact with thermal reservoirs 
at temperatures $T_l$ and $T_r$ : this idea  has been  successfully applied to obtain exact results for   conformal field theories which asymptotically evolve into steady states \cite{1,2,3,4} and to the study of the quantum transport of anyons in one space dimension \cite{pilo,mint0,mint}.

The relativistic counterpart of the Luttinger model is the Thirring model, a model of fermions with  current-current interaction introduced a few years earlier than the Luttinger model  \cite{thirring,wigcar}. The  model is actually the prototype of a large family of two-dimensional models which have been theoretical laboratories to discover non-perturbative  features also shared by realistic four-dimensional models (see   \cite{wigcar} for an old but still good review and  \cite{abdalla} for a more recent survey).

The Thirring and the Luttinger models are indeed closely related:  the directions of the momenta of the fermions in the Luttinger model correspond to  the spin degrees of freedom in the Thirring model. 
There are however substantial differences: the Thirring model is local and  covariant and, as such,  has the standard ultraviolet divergences of relativistic  quantum field theory; on the other hand,  fermions in the Luttinger model interact  non-locally and  ultraviolet divergences are absent. The Thirring model is also plagued by the infrared divergences typical of any local and covariant gauge theory \cite{wigstr,str} which arise here from the  peculiarities of spacetime dimension two; on the other hand the Luttinger model is defined on a compact space and therefore infrared regular; of course  it is not Lorentz covariant but this is not a theme in condensed matter physics.

It took some effort \cite{wigcar} to get a correct vacuum  (i.e. zero temperature) solution of the  Thirring model; Johnson \cite{johnson}   was the first to compute the $n$-point functions at zero temperature by using the Ward identities; his solution was completed by Klaiber \cite{klaiber} who exhibited a quantum field operator having precisely Johnson's $n$-point functions. 
Klaiber's explicit operatorial solution  is  written  in terms of the two key building blocks that are   the massless two-dimensional free scalar field and its dual. Klaiber used a non-covariant but positive definite quantization of the building blocks and recovered Lorentz covariance in the last step. The {\em ab initio} fully covariant quantization was given twenty years later in \cite{strocchib}.
The bosonization technique was    successfully applied also to deal with the  massive case \cite{mand,cole}.

On the other hand,  less attention  has been paid to the thermal representations of the Thirring model  \cite{yoko,sachs,das,pilo,kore,rothe} in comparison with the Luttinger liquids; furthermore, the relevant features have been somehow  buried in irrelevant technical complications; examples are the introduction of fictitious chemical potentials to tame the infrared divergences which at a first glance may seem to be worse than {\em in vacuo} or the redundancy  of the so called thermofield formalism. 

Here we provide some new information about the quantization of two-dimensional massless field and its dual field in the  steady states where there are two independent temperatures for the left and right movers. We face the infrared singularities directly, without introducing artificial chemical potentials for the left and right movers  as is done in the literature.  We provide the general setup by using the Krein space formalism  which is the natural setting when quantizing gauge theories in local and covariant gauges \cite{str,sbook}; this constructions allows in particular an intrinsic  construction of the relevant left and right charge operators  which, as in the zero temperature case \cite{strocchib,strocchi},   belong to the Krein extension of the field algebra and are not introduced by hand. 

The charge operators are in turn a crucial ingredient for the bosonization of the free Dirac field in the steady state with two distinct temperatures for the left and right movers. We recover in our construction a formula by Bernard and Doyon \cite{1}. 

We also construct a general class of interacting models characterized by two temperatures. In particular for the Thirring model \cite{wigcar,klaiber} we provide the full solution in the steady state case. The results described here may be of relevance for further discussions of  integrable models of QFT in 1+1 spacetime dimensions and also for 2-D gravity.

\section{Vacuum of the massless field. Left and right movers}

 Let us first recall the general recipe to obtain a large class of (possibly) inequivalent quantizations of a free massive bosonic field and summarize the main features of the zero temperature massless case.
 The starting ingredient of any  field theory is the covariant commutator:
 \begin{equation}
[\phi(x),\phi(y)] = C(x-y) \label{comm}
 \end{equation}
a  distribution that encodes the commutation relations of the field operators at different spacetime events; for linear field theories it is a $c$-number.

Next, the commutation relations (\ref{comm}) needs to be represented by operators in a Hilbert space. For linear fields, any relevant  Hilbert space  structure may be encoded in a two-point function $W(x,y)$ such that 
  \begin{equation}
 W(x,y) -  W(y,x)=C(x,y). \label{commw}
\end{equation}
If we suppose translation invariance,  $W(x-y)$ may be introduced through its  Fourier representation
\begin{equation}
\widetilde W(k)  =\widetilde \ff(k) \widetilde C(k), \label{wf}
 \end{equation}
where $\widetilde \ff(k)$ is a multiplier for the distribution  $\widetilde C(k)$. The  commutation relations (\ref{commw}) now read 
$
 \widetilde W(k) - \widetilde W(-k)=\widetilde C(k)  
$ 
i.e. 
 \begin{equation}
 \widetilde \ff(k)+\widetilde \ff(-k)=1;
 \end{equation}
 the above condition  must hold on shell, i.e. on a neighborhood of the support of the distribution $\widetilde C(k)$. 

Any choice of the weight $\widetilde  \ff (k)$ such that   the distribution (\ref{wf}) is a positive measure endows 
  the Schwartz space  ${\cal S} ({\bf R}^d)$ of smooth and rapidly decreasing test functions \cite{sw} with a positive semi-definite scalar product which expresses the vacuum-to-vacuum quantum mechanical transition amplitudes encoded in the two-point function:
 \begin{eqnarray}
  \langle  \Psi_f , \Psi_g \rangle=  \langle \Omega, \phi(\bar f) \phi(g) \Omega \rangle= \langle  f , g \rangle = \int\bar f(x)  W(x-y) g(y) dx dy. 
\end{eqnarray}

More specifically, the covariant commutator $C_m(x)$ of the massive Klein-Gordon field  is the unique solution of the Cauchy problem 
\begin{equation}
\left\{ \begin{array}{l}
 (\Box + m^2)C_m(x)  = 0,\cr  
C_m(0 ,{\bf x}) = 0, \cr \partial_0 C_m(0 ,{\bf x}) = - i \delta ({\bf x}) ;
 \end{array} \right.
 \end{equation}
in Fourier space\footnote{Conventions about the Fourier transform in spacetime dimension $d$ are as follows: for test functions
\begin{equation}
f(x) = \frac 1{(2\pi)^d} \int e^{ikx} \widetilde f(k)
\end{equation}
and for distributions 
\begin{equation}
W(x) = \frac 1{(2\pi)^d} \int e^{-ikx} \widetilde W(k)
\end{equation}
so that
\begin{equation}
\int W(x) f(x) dx  
= \frac 1{(2\pi)^{d}}   \int  \widetilde W(k)  \widetilde f(k)dk
\end{equation}
} the solution is written:
 \begin{equation}
  \widetilde C_m(k)= 2\pi  \varepsilon (k^0) \delta(k^2-m^2).
  \end{equation}
The  right choice for the Wightman vacuum is the step function  $ \widetilde \ff(k)=\theta(k^0)$ i.e. only positive energy is allowed for the the states in the Hilbert space of the model:
  \begin{equation}
\widetilde W_m (k)  = \theta(k^0) \widetilde C_m(k)= 2\pi\theta(k^0)\delta(k^2-m^2).
 \end{equation}
The massless limit can be taken straightforwardly:
$
\widetilde W_0(k)  = 
2 \pi \,  \theta(k^0)\delta(k^2) .
 \label{w0k}
$
There is however one notable very important exception:  in spacetime dimension $d=2$ the distribution    $\theta(k^0)$ is not a multiplier for $\delta(k^2)$;  
in that case the limit is well-defined  only on test functions vanishing in Fourier space at $k=0$ 
 \begin{eqnarray}
{\cal S} _0({\bf R}^2) =\left\{ \f\in {\cal S} ({\bf R}^2) , \ \ \ \widetilde  \f (0) =\int \f(x) dx = 0\right\}
\end{eqnarray}
and need to be extended ({\em i.e.}  regularized) to general test function  of ${\cal S} ({\bf R}^2)$. 

It is useful for what follows to recall the simple construction that deals with this problem. By introducing the lightcone variables 
$     x^\pm = x^0\pm x^1$ and $ k^\pm = k^0\pm k^1$ so that $\partial_0 = \partial_++\partial_-$ and $\partial_1 = \partial_+-\partial_-$
one would write 
 \begin{equation} 
  \theta(k^0)\delta(k^2) dk^0dk^1
  = \frac{1}{2k^+}  \theta(k^+) \delta(k^-)  dk^+dk^- + \frac{1}{2k^-} \theta(k^-) \delta(k^+)dk^+dk^-. \end{equation} 
The standard regularization of the rhs goes as follows \cite{gelfand,klaiber}. Consider for instance the right-mover (the first term at the rhs) and define its regularization by subtracting the divergent part: 
 \begin{eqnarray}
W_r(\f)
= \pi  \int_{2\kappa}^\infty  \frac{1}{k^+} \widetilde  \f(k^+,0) dk^+ + \pi  \int^{2\kappa}_{0} \frac{1}{k^+}  \Big[\widetilde \f(k^+,0) -\widetilde \f(0,0)\Big]dk^+
\end{eqnarray}
where $\kappa$ is an arbitrary infrared regulator having the dimension of a mass. A similar definition provides the regularized $ W_l(\f)$ and the complete regularized massless two-point function is the sum
$
 W_0(\f)  =  W_r(\f)  +  W_l(\f).
$

To  compute the Fourier antitransform of the above distributions we may treat the exponential $\exp(-ikx)$ as a test function by adding to $x$  an imaginary vector  in the backward tube \cite{sw}; this is possible because of the positivity of the energy spectrum. 
We get
\begin{eqnarray}
&&W_r(x) =  
- \frac 1 {4\pi} { \log \Big({i \mu  (x^--i \epsilon)}\Big)}, \ \ \ \ \ \mu=e^\gamma \kappa \label{zero}
\end{eqnarray}
(the infrared regulator $\mu$ is not to be confused with a chemical potential).  $W_r(z^-)$ is analytic in the lower half-plane of the complex variable $z^-$. 
An identical calculation provides the Wightman function of the left-mover 
  \begin{eqnarray}
W_l(x) = - \frac 1 {4\pi} { \log \Big({i \mu (x^+-i \epsilon)}\Big)}
\end{eqnarray}
which  is analytic in the lower half-plane of the complex variable $z^+$.

 The left and right movers   are not local fields: the commutators 
   \begin{eqnarray}
C_r(x) =
- \frac i {4} {  \varepsilon(x^-)} , \ \ \ \ \ \ C_l(x) 
= - \frac i {4} {  \varepsilon(x^+)}.
 \label{ccrr}  \end{eqnarray}
do not vanish when $x$ is space-like ($x$ is spacelike if $x^+x^-<0$).
On the other hand their sum  is Lorentz invariant and local:
\begin{eqnarray}
W_0(x) = W_l(x^-)+W_r(x^+) = -\frac 1 {4\pi} \log\Big(-\mu^2 x^2+ i \epsilon x^0\Big), \\
C_0(x) 
=- \frac i {4} {  \varepsilon(x^+)}  - \frac i {4} {  \varepsilon(x^-)}
= -\frac i 2 \varepsilon(x^0)\theta (x^2).
\end{eqnarray}
The distribution  $W_0(x)$ actually extends to a maximally analytic function of the Lorentz invariant complex variable $z^2$ with a  cut on the real positive axis:
  \begin{eqnarray}
 W_0(z) = -\frac 1 {4\pi} \log(-\mu^2 z^2). \label{w0z}
   \end{eqnarray}
However there is a price to pay: having enforced Lorentz invariance and locality we have renounced to positive-definiteness i.e. to a straightforward quantum mechanical interpretation of the model. This is an unavoidable feature of gauge quantum field theories. 

%

%

\section{Introducing the thermal states}

Let us now briefly introduce the discussion with the thermal massive case presented in a way that is suitable for an extension to the massless case. The starting point to  construct the two-point function consists in  inserting the Bose-Einstein distribution for $\tilde{n}(k)$ in Eq. (\ref{wf});
in the massive case this choice is alright  since $\widetilde\ff(k)$ is a well-defined multiplier for the commutator $\widetilde C_m(k)$:
  \begin{equation}
\widetilde W_{m\,\beta} (k) 
= \frac  {2\pi \varepsilon (k^0)\delta(k^2-m^2)}{1-e^{-\beta k^0}}.
 \end{equation}
Moreover, 
the following manipulations are perfectly meaningful:
  \begin{eqnarray}
&& W_{m\,\beta} (x) 
=
\frac 1 {2\pi}
\int  e^{-i  kx } \frac {\theta(k^0)\delta(k^2-m^2)} {1-e^{-\beta k^0}} dk + \frac 1 {2\pi}
\int  e^{-i  kx } \frac {e^{\beta k^0} \theta(-k^0)\delta(k^2-m^2)}{1-e^{\beta k^0}} dk 
\cr 
& =  &
\frac 1 {2\pi}
\sum_{n=0}^{\infty} \int  e^{-i  kx - n \beta k^0 }  {\theta(k^0)\delta(k^2-m^2)} dk + \frac 1 {2\pi}
\sum_{n=1}^{\infty} \int  e^{-i  kx+  n \beta k^0  } { \theta(-k^0)\delta(k^2-m^2)}dk 
\cr 
& =  &
\sum_{n=0}^{\infty} W_m(t -i n \beta, \vec x ) + \sum_{n=1}^{\infty} W'_m(t +i n \beta, \vec x ) \label{ser}
 \end{eqnarray}
where
$  
W'_m(z) = W_m(-z)
$. The series at rhs of (\ref{ser}) converges in the sense of distributions.
\vskip 10 pt
Once more the zero mass limit cannot be taken straightforwardly and it is actually trickier than at zero temperature. For example, following (\ref{ser}) one might try to define the thermal massless two-point function as a series constructed in terms of the massless two-point function (\ref{w0z}) as follows:
   \begin{eqnarray}
W_{0\beta}(t,\vec x)  = -\frac 1 {4\pi} \sum_{n=0}^{\infty} \log\Big(-\mu^2 (t -i n \beta -  \vec x)^2\Big)  
-\frac 1 {4\pi} \sum_{n=1}^{\infty} \log\Big(-\mu^2 (t +i n \beta -  \vec x)^2\Big).  
 \label{w0t}
 \end{eqnarray}
Every term entering in this series is well defined because of the maximal analyticity of $W_0$; the series formally  satisfies the KMS periodicity condition at temperature $T=1/\beta$. But, unfortunately, the series does not converge.

\section{Thermal correlators  of the left and right movers}
Let us proceed {\em ab initio}  as in the zero temperature case by  formally defining the thermal two-point  function by its Fourier transform  as follows:
 \begin{equation}
 \widetilde W_{\beta} (k) = \frac {2\pi \varepsilon (k^0)\delta(k^2) } {1-e^{-\beta k^0}}. \label{wT}
 \end{equation}
 Here we avoid the use of a cutoff in the exponential in the form of a fictive chemical potential, as it is done  in the literature \cite{yoko,pilo} but directly face the infrared divergence. 
At a first superficial glance the infrared divergence seems to be worse than in (\ref{w0k}): the above distribution appears to be well-defined only on test function vanishing at least quadratically at $k=0$; we will clarify below why it is not so.

 The left and right moving parts of the rhs of (\ref{wT}) have now two contributions, according with the sign of the energy ($p$ -- positive, $n$ -- negative):
\begin{eqnarray}
\widetilde W_{r \beta} (k) 
= \widetilde W^p_{r \beta} (k) + \widetilde W^n_{r \beta} (k) 
= \frac{ 2\pi  \theta(k^+) \delta(k^-) }{ {k^+} (1-e^{-\frac 12 \beta k^+})}   + \frac{2\pi e^{\frac 12 \beta k^+}  \theta(-k^+) \delta(k^-) }{ {|k^+|} (1-e^{\frac 12 \beta k^+})} , \label{wplus} \\
\widetilde W_{l \beta} (k) 
= \widetilde W^p_{l \beta} (k) + \widetilde W^n_{l \beta} (k) 
= \frac{  2\pi\theta(k^-) \delta(k^+) }{ {k^-} (1-e^{-\frac 12 \beta k^-})}   + \frac{2\pi e^{\frac 12 \beta k^-}  \theta(-k^-) \delta(k^+) }{ {|k^-|} (1-e^{\frac 12 \beta k^+})} .  \label{wminus}
 \end{eqnarray}
One by one, the  distributions at the rhs are  well-defined only on test functions vanishing at least quadratically at $k=0$.

Let us focus for instance on $ W^p_{r \beta}$:   given a general test function $\f$  we may introduce an  infrared regularized distribution as follows:
 \begin{eqnarray}
 W^p_{r \beta}(\f)&=& 
 \pi  \int_{2\kappa}^\infty  \frac{ \widetilde  \f(k^+,0)}{{k^+} (1-e^{-\frac 12 \beta k^+})}  dk^+ + \pi  \int^{2\kappa}_{0} \frac{ \widetilde  \f(k^+,0)-\widetilde \f(0,0) -\partial_{k^+}\widetilde \f(0,0)\,  k^+}{{k^+} (1-e^{-\frac 12 \beta k^+})} dk^+ + \cr &+&  A \, \widetilde  \f(0,0) + B \, \partial_{k^+}\widetilde \f(0,0).\label{reg}
\end{eqnarray}
The constants $A$ and $B$ are arbitrary and can be adjusted at will.
To compute the  $x$-space representation  of $ W_{r \beta}^p$ we set $\widetilde\f(k^+,0) = e^{-\frac 1  2  i k^+ x^-} $,  expand the denominators at the rhs of Eq. (\ref{reg}) and then interchange the integrals and the series:
 \begin{eqnarray}
W_{r \beta}^p(x)& =&  A + \frac{i B x^-}{2} + \frac 1{4\pi} \, \Big( i \kappa  x^--\log (i \kappa e^\gamma  x^-)\Big) + 
\cr &+&
 \frac 1{4\pi}  \sum_{n=1}^\infty \left[\Gamma (0,n \beta  \kappa )-\frac{i x^- (e^{-\beta  \kappa  n}-1)}{\beta  n}-\log\left(
1+  \frac{  i x^-}{  \beta  n}\right)
\right].
\label{R+nn}\end{eqnarray}
The terms  proportional  to $x^-$ are  crucial for the convergence of the series at the rhs; they come precisely from the subtraction of the term proportional to $k^+$ at the rhs of  Eq. (\ref{reg}). 

A similar expression holds for the negative energy part 
with the noticeable difference that here the expansion starts with $n=1$: 
 \begin{eqnarray}
W_{r \beta}^n(x)=  \frac 1 {4\pi}  \sum_{n=1}^\infty \left[\Gamma (0,n \beta  \kappa )+\frac{i x^- (e^{-\beta  \kappa  n}-1)}{\beta  n}-\log\left(
1-  \frac{  i x^-}{  \beta  n}\right)
\right].
\label{R-nn}\end{eqnarray}
All in all
 \begin{eqnarray}
W_{r \beta}(x) = W_{r \beta}^p(x)+W_{r \beta}^n(x)  &=&  A + \frac{i B x^-}{2} +\frac 1{4\pi} \, \Big( i \kappa  x^--\log (i \kappa e^\gamma  x^-)\Big) + \cr &+&  \frac 1{4\pi}  \sum_{n=1}^\infty \left[2 \Gamma (0,n \beta  \kappa )-\log\left(
1+ \left( \frac{   {x^-}}{  \beta  n}\right)^2\right)
\right].
\label{wrr}
\end{eqnarray}
The terms linear in $x^-$  that guarantee the convergence of  the series (\ref{R+nn}) and (\ref{R-nn}) have disappeared at the rhs of Eq. (\ref{wrr}). The remaining contribution   proportional to  $x^-$  is a regular solution of the wave equation that does not depend on the temperature but gives an anomalous contribution  to the commutator. We may remove it by choosing  $B=-\kappa/2\pi$. 
Similarly we might dispose of  all the constant terms in the series but it is wiser not to do so; we simply take  $A=0$. In the end, setting $\beta = \beta_r$,  $\kappa= \kappa_r$ and $\mu_r=\kappa_r e^\gamma $
 \begin{eqnarray}
W_{r \beta_r}(x) 
  &=&- \frac 1{4\pi} \log (i \mu_r   x^-) -  \frac 1{4\pi} \log\left( \frac{q_r\beta_r \sinh (\frac{\pi  x^-}{\beta_r})}{\pi  x^-}\right)
\label{wrrf} = W_{r}(x)+ T_{r \beta_r}(x) \label{4141}
\end{eqnarray}
where
 \begin{equation}
\log q_r =   \sum_{n=1}^\infty 2 \Gamma (0,n\, \beta_r \kappa_r ).
\label{wrr2} 
\end{equation}

The structure of the final result is quite interesting. The first term $W_{r}(x)$ does not depend on the temperature and  is the only one  contributing to the commutator; it coincides with (\ref{zero}) and as such it  reproduces (\ref{ccrr}).  The second term $T_{r \beta_r}(x)$ does depend on the inverse temperature $\beta_r$. It is a  regular symmetric  function of the variable $x^-$ and therefore has zero commutator:
 \begin{eqnarray}T_{r \beta_r}(x)- T_{r \beta_r}(-x)=0.
\end{eqnarray}
In this sense {\em   $T_{r \beta_r}(x)$ may be understood as a classical correction to the quantum zero-temperature two-point function  $W_{r}(x)$.} Note also that $   T_{r \beta_r}$ tends to zero in the sense of distributions when the temperature vanishes.

 The argument of the $ \log$ in the second term   $T_{r \beta_r}(x)$ is an entire function
of $x^-$. It vanishes only if
$x^- \in i\beta(\bZ\setminus \{0\})$.
In particular it is holomorphic and has no zeros in the set  
$\{x^-\ :\ |\Im x^-| <\beta_r,\}$.
This is to be intersected by the domain of the first term, namely
$\{x^-\ :\ i x^- \notin \bR_-\}$ which contains
$\{x^-\ :\ \Im x^- <0\}$.
Hence the whole expression in (\ref{4141}) is analytic in the strip
 \begin{eqnarray}
\{x^-\ :\ -\beta_r < \Im x^- <0\}
\label{anaR}\end{eqnarray}
and has boundary values in the sense of tempered distributions
at the boundary of the strip. 
As a very important  consequence the Wick powers $W_{r \beta_r}(x)^n$ are
well-defined for any integer $n \ge 0$.

An identical construction provides the thermal equilibrium two-point function for the left mover. Of course there is no necessity to take  the same temperature; in the following we will denote by $\beta_l$ the inverse left temperature; similarly we will denote by $\kappa_l$ the left infrared cutoff: 
 \begin{eqnarray}
W_{l \beta_l}(x) 
  =- \frac 1{4\pi} \log (i \mu_l \, x^+)-  \frac 1{4\pi} \log \left(\frac{q_l  \beta_l\sinh (\frac{\pi  x^+}{\beta_l})}{\pi  x^+}\right).
\label{wrlf}
\end{eqnarray}
Equations (\ref{wrrf}) and (\ref{wrlf}) show that  $W_{r\beta_r}$ and $W_{l\beta_l}$ behave in the infrared better than their 
 positive and negative energy parts and 
 only one subtraction is needed to regularize either $W_{r \beta_r}$ or $W_{l \beta_l}$.  This may also be seen 
by summing Eq. (\ref{reg}) with a similar expression that may readily be written  for $W^n_{r \beta_r}(\f)$; the second subtraction goes away and we get a formula that corresponds to the separation at the rhs of Eq. (\ref{4141}):
\begin{eqnarray}
&& W_{r \beta_r}(\f) = W_{r}(\f) +T_{r \beta_r}(\f)= \pi \int_{2\kappa_r}^\infty  \frac{ \widetilde  \f(k)|_{k^-=0}}  {{2 k^+} } dk^+ + \pi \int^{2\kappa_r}_{0} \frac{ [\widetilde  \f(k)- \widetilde  \f(0)]_{k^-=0}  }{2 {k^+} } dk^+ + 
\cr && 
 + \pi\int_{2\kappa_r}^\infty  \frac{  e^{-\frac 12 \beta_r k^+} [\widetilde  \f(k)+\widetilde  \f(-k)]_{k^-=0}}  {{2 k^+} (1-e^{-\frac 12 \beta_r k^+})}dk^+ +
\pi  \int^{2\kappa_r}_{0} \frac{    e^{-\frac 12 \beta_r k^+}[\widetilde  \f(k)+\widetilde \f(-k)-2\widetilde \f(0) ]_{k^-=0}}{2{k^+} (1-e^{-\frac 12 \beta_r k^+})} dk^+.\cr &&   \label{po} \end{eqnarray}
This formula shows that both $W_r$ and $T_{r\beta_r}$ violate the positive-definiteness unless $h\in {\cal S} _0({\bf R}^2)$.

\subsection{The steady state representation of the  massless scalar field and its dual}
At this point we are in position to consider the steady state  two-point functions for the massless field  and its dual
\begin{equation}
\phi (x) = \phi_r (x^-)+ \phi_l (x^+),  \ \  \tilde \phi (x) = \phi_r (x^-)- \phi_l (x^+):
\end{equation}
\begin{eqnarray}
&&  \langle \Omega, \phi(x) \phi(0) \Omega \rangle =  \langle \Omega, \tilde\phi(x) 
  \tilde \phi(0) \Omega \rangle = W_{r \beta_r}(x) +W_{l \beta_l}(x) = 
  \cr && \cr
&=&  - \frac 1{4\pi} \log (- \mu_r\mu_l   x^2) -  \frac 1{4\pi} \log\left( \frac{q_r\beta_r \sinh (\frac{\pi  x^-}{\beta_r})}{\pi  x^-} \frac{q_l  \beta_l\sinh (\frac{\pi  x^+}{\beta_l})}{\pi  x^+}\right),
\\
&& \cr &&  \langle \Omega, \phi(x)
  \tilde \phi(0) \Omega \rangle =  \langle \Omega, \tilde \phi(x)
 \phi(0) \Omega \rangle = W_{r \beta_r}(x) -W_{l \beta_l}(x) =\cr &&\cr
 &=& - \frac 1{4\pi} \log \left(\frac{ \mu_r   x^-}{\mu_l x^+}\right) -  \frac 1{4\pi} \log\left( \frac{q_r\beta_r \, x^+\,  \sinh (\frac{\pi  x^-}{\beta_r})}{q_l  \beta_l\, x^-\, \sinh (\frac{ \pi  x^+}{\beta_l})}    \right).
\end{eqnarray}
Here we denoted by $\Omega$ the cyclic fundamental state in the Fock space constructed out of the two-point function. Because of the lack of positivity the reconstruction requires an additional structure described  below.

The two-point function is translation invariance is preserved  but of course  not Lorentz invariant. The fields $\phi $ and $\tilde \phi $ are local but not relatively local i.e. they do not commute with each other at spacelike separations. 
The reason behind this failure  \cite{wigstr,str} is  the  local Gauss law 
\begin{equation}
\partial_\mu \tilde \phi = \epsilon_{\mu\nu} \partial^\nu \phi,
\end{equation}
\begin{equation}
\epsilon^{\rho\mu}\partial_\mu \tilde \phi = \epsilon^{\rho\mu}\epsilon_{\mu\nu} \partial^\nu \phi=\partial^\rho \phi.
\end{equation}
%
Let us briefly mention the thermal equilibrium case $\beta_l= \beta_r=\beta$: 
\begin{eqnarray}
W_{\beta}(x)  =
-\frac{1}{ 4\pi}\log (-\mu^2 x^2)
-\frac{1}{ 4\pi}\log \left ( {q^2\, \beta^2 \sinh \left ( {\pi x^- \over \beta} \right )
  \sinh \left ( {\pi x^+ \over \beta} \right ) \over
  \pi^2 x^- x^+ } \right ). \label{W0b}
\end{eqnarray}
Here the whole expression in (\ref{W0b}) is analytic in the tube
\begin{eqnarray}
\Big\{(x^-,\ x^+)\ :\ -\beta < \Im x^- <0,\ \ -\beta < \Im x^+ <0\Big\}
\label{ana}
\end{eqnarray}
and has boundary values in the sense of tempered distributions
at the boundary of the tube. The distributions  $W_{\beta}(x)^n$ are
well-defined for all integers $n \ge 0$ i.e.  the Wick powers of the
field whose two-point function is (\ref{W0b}) are well-defined.
Note that (\ref{W0b}) can also be rewritten as
\begin{align}
W_{\beta}(x) & = 
{1\over 4\pi}\log \left ({\pi^2 \over \mu^2q^2 \beta^2 } \right )
-{1\over 4\pi}\log \left ( -\sinh \left ( {\pi x^- \over \beta} \right )
  \sinh \left ( {\pi x^+ \over \beta} \right ) \right ).
\label{W01}
\end{align}
In particular if we set $x^1 = 0$, $x^+ = x^- = t$, we find
\begin{align}
W_{\beta}(t,\ 0) & = 
{1\over 4\pi}\log \left ({\pi^2 \over \mu^2 q^2 \beta^2 } \right )
-{1\over 2\pi}\log \left (- \sinh \left ( {\pi (t -i\epsilon) \over \beta} \right )
\right ).
\label{W02}
\end{align}

\section{Krein-Hilbert spaces of the left and right movers} 
Here we  consider, as usual in quantum field theory,  $W_{r \beta_r}$ and  $W_{l \beta_l}$  as two-point distributions 
 on 
 ${\cal S} ({\bf R}^2)$,  the Schwartz space  of smooth and rapidly decreasing test functions. 
 As such, they define  two inner products on ${\cal S} ({\bf R}^2)$ that must be used to construct the left and right one-particle spaces and the Fock spaces associated to them:
 \begin{eqnarray}
&& \langle  f , g \rangle_{r  } = \int\overline f(x)  W_{r \beta_r}(x-y) g(y) dx dy , \label{ps1}\\
&& \langle  f , g \rangle_{l  } = \int\overline f(x)  W_{l \beta_l}(x-y) g(y) dx dy 
\end{eqnarray}
 (we left the dependence on the temperature implicit at the lhs).

Let us focus on  $W_{r \beta_r}$.
As in the zero temperature case \cite{strocchi}  $W_{r \beta_r}$ is not positive-semidefinite on ${\cal S} ({\bf R}^2)$ but it is  so when restricted on the chargeless subspace $ {\cal S} _0({\bf R}^2)$. 
Indeed, when either $f$ or $g$ belongs to $ {\cal S} _0({\bf R}^2) $ the scalar product (\ref{ps1})  computed by using  Eq. (\ref{po})  with  $\f(k)= \overline{\widetilde  f (k)}   {\widetilde  g(k)}$    simplifies and no subtraction is  needed anymore:
 \begin{equation}
\langle  f , g \rangle_{r } = 
 \pi  \int_{0}^\infty  \frac{ \overline{\widetilde  f (k)}   {\widetilde  g(k)} |_{ k^-=0}   } {2k^+}
dk^+ 
 + \pi \int_{0}^\infty  \frac{  e^{-\frac 12 \beta_r k^+} [\overline{\widetilde  f (k)}   {\widetilde  g(k)}+\overline{\widetilde  f (-k)}   {\widetilde  g(-k)}]_{k^-=0}}  {{2 k^+} (1-e^{-\frac 12 \beta_r k^+})}dk^+
 .\label{form}
\end{equation}
An analogous formula holds for $W_{l \beta_l}$. 
The non-negativity  of  $\langle  f , f \rangle_{r }$   when $f$  belongs to  $ {\cal S} _0({\bf R}^2) $
is self-evident.

Consider now two general test functions $f,g\in {\cal S} ({\bf R}^2) $. A simplification occurs by choosing a real  test function $\widetilde \chi_r \in {\cal S} ({\bf R}^2)$ such that 
\begin{eqnarray}
&& \langle  \chi_r ,  \chi_r  \rangle_{r } = 0, \ \ \ \ \widetilde  \chi_r  (0) = 1. 
\label{cond}
\end{eqnarray}
It easy to convince oneself that a function with the above properties exists by considering for instance 
$
 \chi_\alpha  = \exp[ -\frac 12 \alpha ({k^+}^2 +{k^-}^2)]
$:
whatever be
the values of the inverse temperature $\beta_r$ and the infrared regulator $\kappa_r$ the expression $\langle  \chi_\alpha ,  \chi_\alpha  \rangle_{r }$ is positive for $\alpha$ sufficiently close to zero and negative for $\alpha$ sufficiently large. Thus, there exists an intermediate value $\alpha_r$ depending on $\beta_r$ and $\kappa_r$ such that the conditions (\ref{cond}) are satisfied.

Given general $f$ and $g$ let  us introduce their chargeless projections 
\begin{eqnarray}
f_0=  f - \widetilde f (0) \chi_r, \ \ \  
g_0=  g - \widetilde g (0) \chi_r,
\end{eqnarray}
so that the pseudo-scalar product (\ref{ps1}) may be rewritten as follows:
\begin{eqnarray}
\langle  f , g \rangle_{r } =  \langle  f _0, g_0 \rangle_{r } +  \overline{\widetilde  f(0)} \langle \chi_r  , g _0\rangle_{r }+
{\widetilde  g(0)} \langle  f _0, \chi_r \rangle_{r };
\end{eqnarray}
in the above formula 
$
\langle   \chi_r, g_0 \rangle_{r \beta_r} $ and $
\langle  f_0,  \chi_r  \rangle_{r \beta_r}  $
are computed by using  Eq. (\ref{form}).
Following step by step the construction displayed in \cite{strocchi} one introduces a Krein-Hilbert majorant topology defined by the scalar product  
\begin{eqnarray}
( f , g )_{r } =  \langle  f _0, g_0 \rangle_{r } +  \overline{\widetilde  f(0)} {\widetilde  g(0)}  + \langle  f _0, \chi_r \rangle_{r } \langle \chi_r  , g _0\rangle_{r }
\label{norm}
\end{eqnarray}
and constructs the one-particle Krein-Hilbert space  $H^{(1)}_{\beta_r}$ of the right mover by completing $ {\cal S}({\bf R}^2) $ w.r.t. the Hilbertian norm (\ref{norm}). 

There exist a special translation invariant vector  
 $v_r\in H^{(1)}_{r\beta_r}$ such that 
\begin{eqnarray}
\langle   \chi_r, f  \rangle_{r } = (v_r,f)_{r }.
\end{eqnarray} 
$v_r$ has Hilbert norm equal to one and, at the same time,  zero expectation value:
\begin{eqnarray}
( v_r, v_r )_{r } = 1 , \ \   \langle v_r, v_r  \rangle_{r } = 0. 
\end{eqnarray}
Actually,  $v_r$ is the strong limit of a sequence of regular test functions converging pointwise to zero\footnote{As an  example consider the sequence $v_n$ constructed as follows:
\begin{eqnarray}
  \chi_n(k) =  \chi_r(k) (1- \phi( n k )), \ \ \ \ v_n(k) = \frac  {\chi_n(k) } { {\langle \chi_n , \chi_r \rangle_r}}
\end{eqnarray}
where $\phi$ is a smooth positive function such that $0\leq \phi \leq 1$, $\phi(0)=1$ and $\phi (k) = 0 $ for $|k|> 1$.
The strong limit $ ||v_n-v_r||_{r } \to 0$ can be shown as follows:
\begin{eqnarray}
&& (v_n ,f  )_{r } =   \langle v_n , f_0 \rangle_{r } +\langle v_n , \chi_r \rangle_{r } \langle \chi_r , f_0 \rangle_{r } = \langle v_n , f_0 \rangle_{r } +\langle \chi_r , f_0 \rangle_{r } \longrightarrow \langle \chi_r , f \rangle_{r \beta_r} ,
\cr
&& (v_n , v_n )_{r } =  1 + \frac { \langle \chi_n , \chi_n \rangle_{r }}  { \langle \chi_n , \chi_r \rangle^2_{r }} \leq 1 + \frac {1}  { \langle \chi_n , \chi_r \rangle_{r }} \longrightarrow 1. \nonumber
\end{eqnarray}}; in this sense $v_r$ can be thought as an infinitely delocalized infrared state \cite{strocchi}. 

Summarizing, $H^{(1)}_{r\beta_r}$  admits the following orthogonal decomposition 
\begin{equation}
H^{(1)}_{r\beta_r} =   L^2_{r\beta_r} ({\bf R}) \oplus v_r\oplus \chi_r\ ; 
\end{equation}
the metric operator $\eta$ representing the two-point function in the Hilbert scalar product 
\begin{equation} \langle \Psi_1,\Psi_2\rangle_r = ( \Psi_1,\eta \Psi_2)_r
\end{equation}
acts like the identity operator when restricted to  $L^2_{r\beta_r}$ and interchanges $\chi_r$ and $v_r$:
\begin{equation}
  \eta=  \left(
\begin{array}{ccc}
 {\bf I}_{L^2} & 0 & 0 \\
 0 & 0 & 1 \\
 0 & 1 & 0 \\
\end{array}
\right), \ \ \ \eta^2={\bf I}.
\end{equation}
The Fock-Segal quantization procedure  may  then be used to obtain the symmetric Fock space of the right mover ${\cal F}({H^{(1)}_{r\beta_r}})$. Similarly one constructs the symmetric Fock space of the left mover ${\cal F}({H^{(1)}_{l\beta_l}})$ and the tensor product  ${\cal F}({H^{(1)}_{l\beta_l}})\otimes {\cal F}({H^{(1)}_{r\beta_r}})$. 
Finally the field operators 
\begin{eqnarray}
\phi_l(f) = \phi^{(+)}_l(f)+\phi^{(-)}_l(f),\ \ \ 
\phi_r(f) = \phi^{(+)}_r(f)+\phi^{(-)}_r(f),
\end{eqnarray} 
operate on the corresponding factors as follows:
\begin{eqnarray}
(\phi^{(+)}_r(f) \Psi_r )^n (x_1,\ldots x_n) &=&  \frac 1 {\sqrt n} \sum_{j=1}^n f(x_j)\Psi_r^{n-1} (x_1,\ldots,\widehat{x}_j,\ldots, x_n), \\
(\phi^{(-)}_r(f) \Psi_r )^n (x_1,\ldots x_n) &=&   \sqrt {n+1} \ \int \overline f(x)W_{r\beta_r}(x,x')  \Psi_r^{n+1} (x' , x_1,\ldots, x_n)dx dx' . \cr &&
\end{eqnarray} 
The above standard formulae may be used to extend the field algebra to include operators such as $\phi_r(f)$ with $f \in H^{(1)}_{r\beta_r}$ and $\phi_l(f)$ with $f \in H^{(1)}_{l\beta_l}$. Particularly important in the following will be the role of the fields $\phi_r^{(\pm )}(v_r)$ and $\phi_l^{(\pm)}(v_l)$;
a simple calculation shows that
\begin{eqnarray}
  [\phi_r^{(-)}(v_r),\phi_r(f) ]=[\phi_r^{(-)}(v_r),\phi^{(+)}_r(f) ] = \langle   v_r, f  \rangle_{r } = (\chi_r,f)_{r } = \widetilde f(0) 
\end{eqnarray} 
so that 
\begin{eqnarray}
 [\phi_r^{(\pm )}(v_r),\phi_r(x) ]=   [ \phi_l^{(\pm )}(v_l),\phi_l(x) ]=  \mp 1.
\label{comv}
\end{eqnarray}

\section{Thermal Fermion Bosonization (two temperatures)}
Here we consider the field algebra ${\cal F}$ generated by the fields $\phi_l$, $\phi_r$ and  their Wick-ordered exponentials $:\exp(z \phi_l):$ and $:\exp(z \phi_r):$\ .  One way to define the Wick exponentials is to ensure the strong convergence of the series 
\begin{equation}  \sum_{n=0}^{\infty}\frac {z^n}{n!} :\phi^n_{l}: (f)  \    , \ \ \ \  \sum_{n=0}^{\infty} \frac {z^n}{n!} :\phi^n_{r}: (f)\ ,\end{equation}
in the Fock-Krein space that we  constructed. The infrared behaviour of the norm associated to the scalar product (\ref{norm}) shows that the above series cannot converge for arbitrary tempered test functions in ${\cal S}({\bf R}^2)$. In the zero temperature case   the Wick-exponentials are Jaffe-type  fields  \cite{jaffe,pierotti,strocchib};  a similar restriction works also in our case but we do not further dwell on this point here and  we limit  ourselves to test functions having compact support in $x$-space where the proof of the strong convergence  of the above series has no substantial difficulty. We do not reproduce the proof here.

Let us consider now the left and right global (or rigid) gauge transformations 
\begin{eqnarray}
&& \gamma_{\lambda}: \ \ \    \phi_l(x^+) \to \phi_l(x^+) + \lambda_l, \\  && \gamma_{\lambda}: \ \ \ \phi_r(x^-) \to \phi_r(x^-) + \lambda_r,  \label{gauge}
\end{eqnarray}
which extend to the Wick exponentials as follows:
\begin{eqnarray}
&&\gamma_{\lambda}: \ \ \ :\exp(z \phi_l):(x^+) \to \ \exp(z\lambda_l) :\exp(z \phi_l):(x^+),   
\\ &&
\gamma_{\lambda}: \ \ \  :\exp(z \phi_r):(x) \to \ \exp(z\lambda_r) :\exp(z \phi_r):(x^-). \label{gauge2}
\end{eqnarray}
These transformations act trivially when the field operators are smeared with chargeless test functions (i.e. such that $\widetilde f(0)=\int f(x) dx =0$).
More generally we may consider the local gauge transformations 
\begin{eqnarray}
&& \gamma_{f}: \ \ \    \phi_l(x^+) \to \phi_l(x^+) + f_l(x^+) , \\  && \gamma_{f}: \ \ \ \phi_r(x^-) \to \phi_r(x^-) + f_r(x_-), \label{gauge3}
\end{eqnarray}
which also extend to the Wick exponentials  under suitable technical conditions on the growth of the functions $f_l(x^+)$ and $f_r(x^-)$ at infinity (which we do not specify here). 
The Hilbert-Krein construction allows for the  existence of  charge operators 
\begin{eqnarray}
Q_l= \frac { i \phi_l^{(+ )}(v_l)-  i \phi_l^{(- )}(v_l) }2, \ \ \ \ Q_r= \frac {i  \phi_r^{(+ )}(v_r)- i  \phi_r^{(- )}(v_r)}2;
\end{eqnarray} 
that implement  the left and right gauge transformations: Eq. (\ref{comv}) indeed is equivalent to   
\begin{eqnarray}
&& \frac d {d \lambda_l }\gamma_{\lambda}( \phi_l(x) )  = i [ Q_l,\phi_l(x)] = 1, \ \ \  \frac d {d \lambda_r }\gamma_{\lambda}( \phi_r(x) )  = i [ Q_r,\phi_r(x)] = 1. \label{gaugeimp}
\end{eqnarray}
The charge operators may be used to introduce  the dressed Wick exponentials:
\begin{eqnarray}
&& \varphi_1 (x)=  e^{  i a  Q_r } : \exp{ z \phi_l }:(x) \, , \\
&& \varphi_2 (x) = e^{  i b  Q_l}  : \exp {z \phi_r }: (x)\, . 
\end{eqnarray} 
Part of what follows goes as in the zero temperature case \cite{strocchib}. Indeed,  set aside  the infrared convergence problems mentioned above,  the possibility to define Wick powers and Wick exponentials of the left and right movers depends only on the ultraviolet behaviour of their correlation functions  and at short distances   the two temperatures play no role. In particular the values of the constants $a,b$ and $z$ that give fermionic anticommutation rules are  the same as in the zero temperature. We briefly sketch the construction for the reader's convenience:  since
\begin{eqnarray}
\varphi_1 (x) \varphi_1 (y)
&=&    
\varphi_1 (y) \varphi_1 (x) \exp\left[ -\frac{i z^2}4 \sgn (x^+-y^+)\right],\\
\varphi_1 (x) \varphi^*_1 (y)
&= &   
\varphi^*_1 (y) \varphi_1 (x) \exp\left[ -\frac{i |z|^2}4 \sgn (x^+-y^+)\right],
\end{eqnarray}
the fields anticommute at spacelike intervals  if there hold the necessary conditions 
$
{ z^2} = \pm 4\pi
$
and
$
{ |z|^2} =  4\pi 
$ 
. Also, the identities
\begin{eqnarray}
\varphi_1 (x) \varphi_2 (y)
&=& \exp[- (b-a) z]  \varphi_2 (y) \varphi_1 (x)\, ,
\\
\varphi_1 (x) \varphi^*_2 (y)&=& 
\exp[( bz +a\overline z) ] \varphi^*_2 (y) \varphi_1 (x)\, ,
\end{eqnarray}
require 
$
 z(b-a) = \pm i \pi 
$
and 
$
 b z +  a \overline z  = \pm i \pi $. 
All in all we get
\begin{eqnarray}
z = 2 i \sqrt {\pi}, \ \ \ \ \ b = a \pm \frac 12  \sqrt {\pi}.
\end{eqnarray}
Let us therefore introduce the fields 
\begin{eqnarray}
&&\psi_1 (x)= \psi_l (x^+)= A \exp  \left(  \frac {\sqrt \pi} {4 i} Q_r \right)  : \exp { \, 2 i \sqrt {\pi} \,  \phi_l }:(x)\, , \label{fermions1} \\
&&\psi_2 (x)= \psi_r (x^-)= B \exp    \left(  \frac {i \sqrt \pi} 4  Q_l \right)   : \exp{\,2 i \sqrt {\pi} \, \phi_r }:(x) \,  , \label{fermions2}
\end{eqnarray} 
and compute their anticommutators:
\begin{eqnarray}
\psi_1(x) \psi_1^*(y) + \psi_1^*(y) \psi_1(x) = {|A|^2\pi (x^+-y^+) \over q_l 
\mu_l \beta_l  \sinh\left ({\pi (x^+-y^+) \over \beta_l}\right )}
:e^{2i\sqrt {\pi}\phi_l(x)}e^{-2i\sqrt {\pi}\phi_l(y)}:
\times \cr \times \left [ {1 \over i(x^+-y^+)+\veps} + {1 \over i(y^+-x^+)+\veps}
\right ].
\end{eqnarray}
The last factor equals $2\pi\delta(x^+-y^+)$. At $(x^+-y^+) =0$ the Wick product becomes equal to 1 and the first factor is regular:
\beq
\psi_1(x) \psi_1^*(y) + \psi_1^*(y) \psi_1(x) =
{2\pi |A|^2 \over q_l\mu_l   } \delta(x^+-y^+)\ .
\endq
Similarly
\begin{eqnarray}
\psi_2(x) \psi_2^*(y) + \psi_2^*(y) \psi_2(x) &=&
{2\pi|B|^2 \over q_r  \mu_r } \delta(x^--y^-)
\end{eqnarray}
while all the remaining anticommutators are zero. By choosing  
\begin{equation}
A = \sqrt { q_l \mu_l \over 2\pi },\ \ \ \ \ B = \sqrt { q_r \mu_r \over 2\pi },
\end{equation}
the above construction reproduces the { canonical commutation relations} of a massless free  Dirac field. 

The Dirac equation\footnote{Gamma matrices are as follows:
    \begin{eqnarray}
\gamma^0=\left(
\begin{array}{cc}
 0 & 1 \\
 1 & 0 \\
\end{array}
\right), \ \ \ \gamma^1=\left(
\begin{array}{rr}
 0 & 1 \\
 -1 & 0 \\
\end{array}
\right), \ \ \ \gamma^5=\gamma^0\gamma^1=\left(
\begin{array}{rr}
 -1 & 0 \\
 0 & 1 \\
\end{array}
\right).
    \end{eqnarray}} is satisfied by construction:
\begin{eqnarray}
  \gamma^\mu \partial_\mu \psi= 
\left[
\begin{array}{l}
\partial_+\psi_r (x^-) \\
\partial_- \psi_l(x^+) \\
\end{array}
\right] = 0.
    \end{eqnarray}
The two-point  expectation value in the fundamental state 
\begin{eqnarray}
    \langle \Omega, \psi(x) \overline{\psi}(y)\Omega\rangle
&=& \left(
\begin{array}{cc}
 0 & |A|^2 e^{4\pi W_{l\beta_l}(x-y)} \\
|B|^2 e^{4\pi W_{r\beta_l}(x-y)} & 0 \\
\end{array}
\right) = \cr &=&\left(
\begin{array}{cc}
 0 & \frac{1}{2i    \beta_l\sinh \frac{\pi(  x^+-y^+)}{\beta_l}} \\
\frac{1}{2i    \beta_r\sinh \frac{\pi  (x^--y^-)}{\beta_r}} & 0 \\
\end{array}
\right)   \end{eqnarray}
 does not depend on the cutoffs  but only on the left and right temperatures, as it must be,  because a massless Dirac field  in 1+1 spacetime dimension is not infrared singular (as opposed to the scalar field $\phi$).

On the other hand the $n$-point  expectation values {\em do not} coincide with those of a free Dirac spinor field:  many of them  should be zero but they are not. As in the zero temperature case, one may impose by hand {\em ad hoc} selection rules to kill all the unwanted contributions \cite{wigcar} or introduce an extra free fermion \cite{klaiber}. Fortunately, in the present framework these steps are not necessary: in the Hilbert-Krein-Fock space displayed in the previous section there exist special states, called {\em fermionic vacua} \cite{strocchib} (see Appendix \ref{vacua}) which automatically implement the necessary  selection rules. 

In particular, for the free Dirac fields (\ref{fermions1}) and (\ref{fermions2}) 
the fermionic states  we are referring to are labelled by two angles $\theta_l$ and $\theta_r$ as follows :
\begin{equation}
\Omega_{\theta_l\theta_r}^{free} =\frac 1 {4 \pi}  \int^{\sqrt{\pi}}_{-\sqrt{\pi}}d\lambda_r
 \int^{\sqrt{\pi}}_{-\sqrt{\pi}}d\lambda_l
e^{i(\lambda_r+\theta_r)Q_r+i(\lambda_l+\theta_l)Q_l}  \  \Omega.
\end{equation}
The two-point expectation value in the state $\Omega_{\theta_l\theta_r}^{free}$ coincides with the  expectation value in the fundamental state
\begin{eqnarray}
    \langle\Omega_{\theta_l\theta_r}^{free}, \psi(x) \overline{\psi}(y)\Omega_{\theta_l\theta_r}^{free}\rangle = 
\langle \Omega, \psi(x) \overline{\psi}(y)\Omega\rangle  \end{eqnarray}
and the truncated $n$-point functions in the state $\Omega_{\theta_l\theta_r}^{free}$ vanish. The steady state fermion bosonization is thus achieved.

As is well-known \cite{dashen,sugawara} the classical currents
\begin{eqnarray}
   &&j^{\mu}(x) =\bar \psi(x)\gamma^\mu  \psi(x),
\\
   &&\tilde j^{\mu}(x) =\bar \psi(x)\gamma^5\gamma^\mu  \psi(x) = \epsilon^{\mu\nu}j_\nu
\end{eqnarray}
require an ultraviolet regularization at the quantum level. 
A renormalized current may be obtained as follows
\begin{eqnarray}
 j_{+}(x)&= & j_{0}(x)+j_{1}(x) = 2 \lim_{\epsilon\to 0} [ \psi_1^{*}(x+\epsilon)  \psi_1(x) -  \langle \psi_1^{*}(x+\epsilon)  \psi_1(x) \rangle] = 
 \cr 
&=&  -  2
 \sqrt{\pi} \  \epsilon^+ \partial_+ \phi_l(x^+)    \frac{1}{   \beta_l\sinh \frac{\pi  \epsilon^+}{\beta_l}}\to   -  
\frac {2}{ \sqrt{\pi} }\   \partial_+ \phi_l(x^+). 
\end{eqnarray}
Similarly
\begin{eqnarray}
   j_{-}(x)&=& -
\frac {2}{ \sqrt{\pi} }\   \partial_- \phi_r(x^-) .  
   \end{eqnarray}
There follow the  commutation relations \cite{dashen,sugawara,della} 
\begin{eqnarray}
[ j_{+}(x),j_+(y) ]= 
 {2 i \xi }    \delta'(x^+-y^+),\ \ \ \       
[ j_{+}(x),j_+(y) ]= 
{2 i \xi }   \delta'(x^--y^-), \ \ \ \ 
\xi =\frac {2 } {\pi}. 
   \end{eqnarray}
An elementary computation gives (see Eq. (\ref{wrlf}))
\begin{eqnarray}
&&  \langle \Omega, j_+(x) j_+(0) \Omega \rangle = 
-\frac 4 {\pi} \partial_+^2 W_{l \beta_l}(x^+) = 
-  \frac{1}{\beta_l^2\sinh^2 (\frac{\pi  x^+}{\beta_l})}.
\end{eqnarray}
By subtracting the vacuum (i.e. zero temperature) expectation value we get the normal ordered  two-point  expectation value of the current 
\begin{eqnarray}
&&  \langle \Omega,: j_+(x) j_+(0) : \Omega \rangle  =
 -\frac 4 {\pi} \partial_+^2 (W_{l \beta_l}(x^+) - W_{l }(x^+)) = \cr && =   \frac 1{\pi^2}\partial_+^2 \log \left(\frac{q_l  \beta_l\sinh (\frac{\pi  x^+}{\beta_l})}{\pi  x^+}\right)=-  \frac{1}{\beta_l^2\sinh^2 (\frac{\pi  x^+}{\beta_l})}+ 
\frac{1}{\pi ^2 {x^+}^2}.\end{eqnarray}
We can now take the limit where the two points coincide and see that there are persistent squared currents
\begin{eqnarray}
 \langle \Omega,: j^2_+(x)  : \Omega \rangle  =\frac{1}{3\beta_l^2} \ \ \ \langle \Omega,: j^2_-(x)  : \Omega \rangle  =\frac{1}{3\beta_r^2}
 \end{eqnarray}
and non-vanishing expectation values of the energy momentum tensor \cite{dashen,sugawara,della} 
\begin{eqnarray}
&&  \Theta_{\mu\nu} (x)  = \frac 1{2\xi} :\left( 2 j_{\mu} j_{\nu}-g_{\mu\nu} j^{\alpha}j_{\alpha}\right): (x),
 \end{eqnarray}
\begin{eqnarray}
  \langle \Omega,: \Theta_+(x)  : \Omega \rangle  =\frac{\pi}{12\beta_l^2}, \ \ \ \   \langle \Omega,: \Theta_-(x)  : \Omega \rangle  =\frac{\pi}{12\beta_r^2}.
 \end{eqnarray}
By considering the the total energy flow we are able to  reproduce a formula by Bernard and Doyon \cite{1}  in the special case where the central charge is equal to one:
\begin{eqnarray}
&&  \langle \Omega,: (\Theta_+(x)  - \Theta_-(x) ) : \Omega \rangle  =\frac{\pi}{12\beta_l^2}-\frac{\pi}{12\beta_r^2}.
 \end{eqnarray}

\section{Interacting fermions: the steady-state Thirring model}

Here we construct the full operator solution of the Thirring model in the general  steady state with a left and right temperature.
Following \cite{strocchi}, let us at first introduce generic  non-free  fermion operators in terms of five real parameters $a,b,c,d,\sigma$ as follows:
\begin{eqnarray}
&&\psi_1 (x)= A \exp  \left(  \frac {\sigma \pi} {4 i {c}} Q_l  +\frac {\sigma \pi} {4 i {d}} Q_r \right)  : \exp { ( i {a}  \,  \phi_l+ i {b}  \,  \phi_r )}:(x)\, ,  \label{1deff} \\ 
&&\psi_2 (x)= B \exp    \left(  \frac {i\sigma  \pi}{4 {a} }  Q_l +\frac {i\sigma \pi} {4 {b}} Q_r  \right)   : \exp { ( i {c}  \,  \phi_l+ i {d}  \,  \phi_r )}:(x)  \,  .  \label{1defg}
\end{eqnarray} 
 Their commutation relations are easily computed:
\begin{eqnarray}
\psi_1 (x) \psi_1 (y)
%
%
&=&    
\psi_1 (y) \psi_1 (x) \ e^{ \frac{i {a}^2}4 \sgn (x^+-y^+)+ \frac{i {b}^2}4 \sgn (x^--y^-)},\\
 \psi_2 (x) \psi_2 (y)&=& \psi_2 (y) \psi_2 (x)\  e^{\frac{i {c}^2}4 \sgn (x^+-y^+)+ \frac{i {d}^2}4 \sgn (x^--y^-)},
\\
\psi_1 (x) \psi_2 (y)
%
%
 %
 &=&  \psi_2 (y) \psi_1 (x)  e^{ -  {i \sigma\pi} }  e^{ \frac{i {a}{c}}4 \sgn (x^+-y^+)+ \frac{i {b}{d}}4 \sgn (x^--y^-)} . 
\end{eqnarray}
 At  spacelike separated events $x$ and $y$ the fields should anticommute (see \cite{mint} for the anyonic case); for that to happen the following   conditions must hold:
\begin{equation*}
{a}^2-{b}^2 = 4(2n+1)\pi, \ \ \ {c}^2-{d}^2 = 4(2m+1)\pi,
\end{equation*}
\begin{equation}
{a}{c}-{b} {d} = 4k \pi,  \ \ \   \sigma = \frac 12 [1+(-1)^k], \label{zuf}
\end{equation}
where $n$, $m$ and $k$ are integers. These condition  are algebraic, do not  and cannot depend on the temperatures. 
As regards the currents,  at first order in $\epsilon$ we have the following 
\begin{eqnarray}&& \psi_1^{*}(x+\epsilon)  \psi_1(x) -  \langle \psi_1^{*}(x+\epsilon)  \psi_1(x) \rangle 
\simeq  -  i A^2
  \ (  {a}  \ \epsilon^+ \partial_+ \phi_l(x^+)  +  {b} \ \epsilon^- \partial_- \phi_r(x^-)  ) \times \cr    & \times&  \left(\frac{i \mu_l  q_l \,   {\beta}_l\sinh (\frac{\pi  \epsilon^+}{{\beta}_l})}{\pi  }\right)^{-\frac{{a}^2}{4\pi}}    \left(\frac{i \mu_r  q_r \, {\beta}_r\sinh (\frac{\pi  \epsilon^-}{{\beta}_r})}{\pi  }\right)^{-\frac{{b}^2}{4\pi}}.
  \end{eqnarray}    
An obvious  multiplicative renormalization produces a  quantum current which explicitly breaks Lorentz covariance:
\begin{eqnarray}
&&  e^{  (4\pi -{a}^2 )W_{l{b}_l} (\epsilon)-  {b}^2 W_{r{\beta}_r} (\epsilon)}  
[ \psi_1^{*}(x+\epsilon)  \psi_1(x) -  \langle \psi_1^{*}(x+\epsilon)  \psi_1(x) \rangle] \to  \cr  && 
 \to  
 - \frac{   {a} }{2 \pi}  \ \partial_+\phi_l(x^+)   - \frac{   {b} }{2 \pi} \ \frac{\epsilon^-}{\epsilon^+} \partial_- \phi_r(x^-) .
   \end{eqnarray}    
This shortcoming  may be corrected along the lines indicated in  \cite{klaiber,strocchi} by introducing the currents $J_+$ and $J_-$ as follows:
\begin{eqnarray}
J_+ &=& 
 \frac{ 2 \psi_1^{*}(x+\epsilon)  \psi_1(x) -  2 \langle \psi_1^{*}(x+\epsilon)  \psi_1(x) \rangle (1 + i g_+\epsilon^+ J_+ + i g_-\epsilon^- J_-) }{  e^{  ({a}^2-4\pi  )W_{l{\beta}_l} (\epsilon)+  {b}^2 W_{r{\beta}_r} (\epsilon)}  } \cr
  &\simeq&  - \frac{ 1 }{2\pi}  \left(  {a}  \ \partial_+\phi_l(x^+)  +  {b} \ \frac{\epsilon^-}{\epsilon^+} \partial_- \phi_r(x^-)  \right) -  \frac{  g_+  J_+ }  { 2 \pi    } -  \frac{  g_-\epsilon^- J_-}  { 2 \pi  \epsilon^+  },   \end{eqnarray}    
\begin{eqnarray}
J_- &=& \frac{[ \psi_2^{*}(x+\epsilon)  \psi_2(x) -  \langle \psi_2^{*}(x+\epsilon)  \psi_2(x) \rangle (1 + g_+\epsilon^+ J_+ + g_-\epsilon^- J_-)] }{  e^{  {c}^2 W_{l{\beta}_l} (\epsilon)+ ({d}^2-4\pi )^2 W_{r{\beta}_r} (\epsilon)}  } \cr  &&\simeq   - \frac{ 1}{2\pi}  \left(  {c}  \frac{\epsilon^+}{\epsilon^-}  \ \partial_+\phi_l(x^+)  +  {d} \, \partial_- \phi_r(x^-)  \right) -  \frac{  g_+\epsilon^+ J_+}  {2 \pi  \epsilon^-}-  \frac{   g_-J_-}  {2 \pi }.
 \end{eqnarray} 
      %
     %
 %
     At the algebraic level Lorentz covariance is maintained if 
\begin{equation}
\frac{  g_+} {  \pi }  = \frac{  {c}}   { {a} -{c}},   \ \ \ \ \frac{  g_-} {  \pi }  = \frac{  {b}}   { {d} -{b}\,} , 
\end{equation}
which imply     
\begin{eqnarray}
J_+ =
 - \frac{ 1 }{\pi}   ({a} -{c}) \ \partial_+\phi_l(x^+),   \ \ \ \
J_- =
 - \frac{ 1}{\pi} ({d} -{b})\, \partial_- \phi_r(x^-),
   \end{eqnarray}  
%
  %
  %
 or equivalently 
 \begin{eqnarray}
&& J_\mu =  
 - \frac{ 1 }{4 \pi}   ({a} -{c}+{d}-{b}) \ \partial_\mu \phi    - \frac{ 1}{4\pi} ({c}-{a}+{d} -{b})\, \partial_\mu \tilde\phi, 
\\
&& \widetilde J_\mu =  \epsilon_{\mu\nu} J^{\nu}
 =
 - \frac{ 1 }{4 \pi}   ({a} -{c}+{d}-{b}) \  \partial_\mu \tilde  \phi    - \frac{ 1}{4\pi} ({c}-{a}+{d} -{b})\,  \partial_\mu \phi. 
   \end{eqnarray} 
 The condition 
\begin{equation}
    {d}-{b} = {a} -{c} \label{convec}
\end{equation} 
is imposed to preserve the vector and pseudovector characters of the currents $J_\mu$ and respectively $\widetilde J_\mu$.
Of course in the steady-steate representation  Lorentz symmetry is not implemented; the correlation functions depend on the two temperatures and only translation symmetry is unbroken. 

Let us  focus now on the field equation:  
\begin{eqnarray}
 i  {\gamma}^\mu \partial_\mu \psi
= - \left(
\begin{array}{lr}
 0 & c\,  \partial_+\phi_l  \\
 {b}\, 
\partial_- \phi_r  & 0 \\
\end{array}
\right) \left[
\begin{array}{l}
\psi_1 (x) \\
\psi_2(x) \\
\end{array}
\right].
\end{eqnarray}
After some manipulations and the necessary ultraviolet regularization  it may be rewritten as follows:
 \begin{eqnarray}
 i  \gamma^\mu \partial_\mu \psi  &=&  - \frac 12( b+ c)
\ \gamma^\mu :\partial_\mu \phi \ \psi :(x)  - \frac 12( b-c)
\ \gamma^\mu  :\partial_\mu \tilde \phi \ \psi :(x) \\
 &=&  \pi\,  \frac { b+ c}{{a}-c}
\ \gamma^\mu :J_\mu  \ \psi :(x) + \pi\,  \frac { b- c}{{a}-c}
\ \gamma^\mu  : \tilde J_\mu \ \psi :(x).
\end{eqnarray}
 The special choice $b= c$ (which implies ${a}=d$) corresponds to the Thirring model:
\begin{eqnarray}
i  \gamma^\mu \partial_\mu \psi(x)=\,  \frac { 2 \pi b }{{a}-b}  :\gamma^\mu J_\mu \, \psi : (x),
    \end{eqnarray}
%

The operator solution is then written as follows
\begin{eqnarray}
&&\psi_1 (x)= A \exp  \left(  \frac { \pi} {4 i b } Q_l  +\frac { \pi} {4 i {a}} Q_r \right)  : \exp { ( i {a}  \,  \phi_l+ i b  \,  \phi_r )}:(x)\, ,  \label{deff} \\ 
&&\psi_2 (x)= B \exp    \left(  \frac {i  \pi}{4 {a} }  Q_l +\frac {i \pi} {4 b} Q_r  \right)   : \exp { ( i b  \,  \phi_l+ i {a}  \,  \phi_r )}:(x)  \,  ,  \label{defg}\\ 
   \end{eqnarray}  
 with ${a}^2-b^2 = 4(2n+1)\pi$. Note that in this case $\sigma=1$ and the occurrence of the infrared operators in Eqs. (\ref{deff}) and (\ref{defg})  is unavoidable; they  play the role of Coleman's spurions \cite{cole} but here, as in \cite{strocchib}, they emerge naturally in the construction. Here 

The global gauge transformations (\ref{gauge}) now read
\begin{eqnarray}
&&\gamma_{\lambda}: \ \ \ \psi_1(x)  \to \ \exp(i a \lambda_l+i b \lambda_r) \psi_1(x),   
\\ &&\gamma_{\lambda}: \ \ \ \psi_2(x)  \to \ \exp(i b \lambda_l+i a \lambda_r) \psi_2(x).  
\end{eqnarray}
Define 
$$ \Omega_{\lambda\tilde\lambda}= e^{i\lambda(Q_l+ Q_r)}  e^{i\tilde \lambda(Q_l- Q_r)}  \  \Omega.$$
Since 
\begin{eqnarray}
&&\langle \Omega_{\lambda\tilde\lambda}, :e^{ i n_1(  a  \phi_l+  b    \phi_r )}:(x_1)\ldots :e^{ i n_k(  a  \phi_l+  b    \phi_r )}:(x_k):e^{ i m_1(  b  \phi_l+  a    \phi_r )}:(y_1)\ldots \times \cr && \times  :e^{ i m_q(  b  \phi_l+  a    \phi_r )}:(y_q) \Omega_{\lambda'\tilde\lambda'}\rangle = \cr  &&\cr &=&
 e^{-\frac i 2 (\lambda_l+\lambda'_l)(a+b)(n_1+\ldots+n_k+m_1+\ldots +m_q)}  e^{-\frac i 2 (\tilde\lambda_l+\tilde\lambda'_l)(a-b)(n_1+\ldots+n_k-m_1+\ldots -m_q)} \times \cr && \times \langle \Omega, :e^{ i n_1(  a  \phi_l+  b    \phi_r )}:(x_1)\ldots :e^{ i n_k(  a  \phi_l+  b    \phi_r )}:(x_k):e^{ i m_1(  b  \phi_l+  a    \phi_r )}:(y_1)\ldots \times \cr && \times  :e^{ i m_q(  b  \phi_l+  a    \phi_r )}:(y_q) \Omega\rangle
\end{eqnarray}
the relevant fermionic vacuum $\Omega_T$ is given by (see Appendix \ref{vacua})
\begin{eqnarray}
\Omega_T =  \frac {a^2-b^2}{16 \pi^2}\int^{\frac{2\pi}{a+b}}_{-\frac{2\pi}{a+b}}d {\lambda}   \int^{\frac{2\pi}{a-b}}_{-\frac{2\pi}{a-b}}d {\tilde \lambda} \ \Omega_{\lambda\tilde\lambda}. \end{eqnarray}
It provides the necessary selection rules. In particular the non zero two-point functions  are
 \begin{eqnarray}
&&  \langle \Omega_{T}, \psi^\dagger_1(x) {\psi_1}(y)\Omega_{T}\rangle=
\left({ 2 i \,  \beta_l\sinh \left(\frac{\pi  (x^+-y^+)}{\beta_l}\right)}{ }\right)^{-\frac{a^2}{4\pi}}  \left({ 2i   \beta_r\sinh \left(\frac{\pi ( x^--y^-)}{\beta_r}\right)}{  }\right)^{-\frac{b^2}{4\pi}} \cr &&
\\ &&
\cr
&&  \langle \Omega_T, \psi^\dagger_2(x) {\psi_2}(y)\Omega_T\rangle=
\left({ 2 i \,  \beta_l\sinh \left(\frac{\pi  (x^+-y^+)}{\beta_l}\right)}{ }\right)^{-\frac{b^2}{4\pi}}  \left({ 2i   \beta_r\sinh \left(\frac{\pi  (x^--y^+)}{\beta_r}\right)}{  }\right)^{-\frac{a^2}{4\pi}} \cr &&
  \end{eqnarray}
where 
 \begin{eqnarray}
  A^2    =  \left(\frac{  q_l \,  \mu_l  \, }{2 \pi  }\right)^{\frac{a^2}{4\pi}}  \left(\frac{  q_r \,  \mu_r \,  \, }{2\pi  }\right)^{\frac{b^2}{4\pi}} \, , \ \ \  B^2    =  \left(\frac{  q_l \,  \mu_l \, , }{2 \pi  }\right)^{\frac{b^2}{4\pi}}  \left(\frac{  q_r \,  \mu_r \, \, }{2\pi  }\right)^{\frac{a^2}{4\pi}} .
  \end{eqnarray}
We see that in the interacting fields the left and right components are no more separated; the correlation functions are products of factors which are functions of $x^+$ and $x^-$.  The two inverse temperatures appears in every correlator.
Finally, proceeding as before we get
 \begin{eqnarray} \langle \Omega,: J^2_+(x)  : \Omega \rangle  =\frac{(a-b)^2}{4\pi}\frac{1}{3\beta_l^2} \ \ \ \langle \Omega,: J^2_-(x)  : \Omega \rangle  =\frac{(a-b)^2}{4\pi}\frac{1}{3\beta_r^2} \end{eqnarray}
But, since 
\begin{equation}
[ J_{+}(x),J_+(y) ]= 
 {2 i\xi }    \delta'(x^+-y^+),\ \         
[ J_{+}(x),J_+(y) ]= 
{2 i \xi}   \delta'(x^--y^-), \ \   
\xi =\frac {(a-b)^2 } {2\pi}   \end{equation} 
the expectation values of the  energy-momentum tensor  remain the same as before:
\begin{eqnarray}
  \langle \Omega,: \Theta_+(x)  : \Omega \rangle  =   \frac 1 {2\xi} \langle \Omega,: J^2_+(x)  : \Omega \rangle  =\frac{\pi}{12\beta_l^2}, \\
  \langle \Omega,: \Theta_-(x)  : \Omega \rangle = \frac{1}{2\xi}  \langle \Omega,: J^2_-(x)  : \Omega \rangle  =\frac{\pi}{12\beta_r^2}.
 \end{eqnarray}

\section{Conclusions}
Krein space techniques  provide  not-so  well-known and yet  powerful tools  to deal with gauge QFT's;  they are in particular very effective to construct operator solutions  of models of QFT in spacetime dimension two. 

The great advantage of Krein space techniques is that  they are directly related to the infrared behaviour of the  correlation functions of the quantum fields; some of the ingredients that are introduced ad hoc in other constructions such as the infrared operators,  here emerge naturally.  

In this paper we have used the above techniques to construct thermal representations for the massless boson and spinor field. We have treated the left and right degrees of freedom  independently. As an application we have discussed a full operator solution of the Thirring field in the steady state where there are two independent left and right temperatures. It is seen in the construction the importance of the infrared degrees of freedom that naturally emerge in our context. 

The methods and the results discussed in this paper may prove to be useful also in other contexts where the massless two-dimensional field is relevant such as conformal field theory and string theory.
\begin{acknowledgments}
A correspondence with Mihail Mintchev and the suggestions of an anonymous referee are gratefully acknowledged
\end{acknowledgments}
\appendix
\section{Fermionic vacua} \label{vacua}

An element $\Phi$ of the field algebra ${\cal F}$ has left  charge $q_l$ and right charge $q_r$ if 
\begin{equation}
    \gamma^{\lambda_l}_l(\Phi) = e^{i \lambda_l q_l} \Phi , \ \ \  \gamma^{\lambda_r}_r(\Phi) = e^{i \lambda_r q_r} \Phi.
\end{equation}
Examples  are the Wick exponentials
\begin{eqnarray}
&&  :\exp(i q_l \phi_l):(x) \to \ \exp(i\lambda_l q_l) :\exp(iq_l \phi_l):(x)   \\ &&   :\exp(i q_r \phi_r):(x) \to \ \exp(i \lambda_r q_r) :\exp(iq_r \phi_r):(x). \label{gauge2bis}
\end{eqnarray}
Fields having  left  and right charges which are integer multiples of  $q_l$ and $q_r$ constitute a subalgebra of  
${\cal F}$ which we denote ${\cal F}_{q_lq_r}$. The following simple observation is nonetheless crucial: formula (\ref{gaugeimp}) implies that 
\begin{eqnarray}
&&\langle \Omega, \  e^{-i\lambda_l Q_l }\  :e^{i q_l \phi_l}:(x)   \ e^{i\lambda'_l Q_l } \Omega\rangle =\cr &=&
\langle \Omega, \  e^{-\frac 12 \lambda_l \phi^{(-)}_l(v_l) }\  :e^{i q_l \phi_l}:(x)   \  e^{-\frac 12 \lambda'_l \phi^{(+)}_l(v_l)  }\Omega\rangle = e^{-\frac i 2 (\lambda_l+\lambda'_l)q_l}.
\end{eqnarray}
 This promptly generalizes to 
\begin{eqnarray}
&&\langle \Omega, \  e^{-i\lambda_l Q_l }\  :e^{(i  q_l n_1\phi_l)}:(x_1) \ldots   :e^{i  q_l n_k\phi_l}:(x_k)  \ e^{i\lambda'_l Q_l } \Omega\rangle = \cr  &=&
 e^{-\frac i 2 (\lambda_l+\lambda'_l)q_l(n_1+\ldots+n_k)} \langle \Omega, \   :e^{i  q_l n_1\phi_l)}:(x_1) \ldots  :e^{i  q_l n_k\phi_l}:(x_k)  \ \Omega\rangle
\end{eqnarray}
so  that 
\begin{eqnarray}
&&\int^{\frac{2\pi}{q_l}}_{-\frac{2\pi}{q_l}}\langle \Omega, \  e^{-i\lambda_l Q_l }\  :\exp(i  q_l n_1\phi_l):(x_1) \ldots   :\exp(i  q_l n_k\phi_l):(x_k)  \ e^{i\lambda'_l Q_l } \Omega\rangle d\lambda_l= \cr &&= {\frac{4\pi}{q_l}}\delta_{0,n_1+\ldots +n_k}  \langle \Omega, \   :\exp(i  q_l n_1\phi_l):(x_1) \ldots   :\exp(i  q_l n_k\phi_l):(x_k)  \ \Omega\rangle.
\end{eqnarray}
It follows the correlation function  $\langle \Omega^{q_lq_r},\Phi \, \Omega^{q_lq_r} \rangle$ of  an operator $\Phi$ belonging to ${\cal F}_{q_lq_r}$  on the state 
\begin{eqnarray}
\Omega^{q_lq_r} = \frac {q_lq_r}{16\pi^2} \int^{\frac{2\pi}{q_l}}_{-\frac{2\pi}{q_l}}d {\lambda_l}   \int^{\frac{2\pi}{q_r}}_{-\frac{2\pi}{q_r}}d {\lambda_r} e^{i\lambda_l Q_l}  e^{i\lambda_rQ_r } \ \Omega
\end{eqnarray}
 vanishes unless his charges are equal to zero:  $\gamma^{\lambda_l}_l(\Phi)  = \gamma^{\lambda_r}_r(\Phi) =0$. More generally we may introduce two angular parameters and define the the so called $\theta$-vacua states: 
\begin{eqnarray}
\Omega_{\theta_l\theta_r}^{q_lq_r} = \frac {q_lq_r}{16\pi^2} \int^{\frac{2\pi}{q_l}}_{-\frac{2\pi}{q_l}}d {\lambda_l}   \int^{\frac{2\pi}{q_r}}_{-\frac{2\pi}{q_r}}d {\lambda_r} e^{i(\lambda_l +\theta_l)Q_l}  e^{i(\lambda_r+\theta_r)Q_r }\Omega
\end{eqnarray}
such that the correlation functions of fields belonging to ${\cal F}_{q_lq_r}$  vanish unless they are chargeless.


\begin{thebibliography}{99}
\bibitem{tomonaga}  S. Tomonaga, 
 Prog. Theor. Phys. (Kyoto) 5, 544 (1950).


\bibitem{bloch1} F. Bloch, Z. Physik 81, 363?376 (1933). 

\bibitem{bloch2}  F. Bloch, Helv. Pbys. Acta 7, 385 (1934).

\bibitem{luttinger} J. M. Luttinger, J. Math. Phys. 4:1154 (1963).
\bibitem{lieb}{ E. H. Lieb and D. C. Mattis, J. Math. Phys. 6:304 (1965).}

\bibitem{haldane} P. D. M.  Haldane, J. Phys. C: Solid State Phys. 14, 2585 (1981)

\bibitem{hei} R. Heidenreich, R. Seiler, D.A. Uhlenbrock,  J. Stat. Phys. 22, 27? 57 (1980) 
\bibitem{1}
 D. Bernard and B. Doyon, J. Phys. A: Math. Theor. 45, 362001
(2012).
\bibitem{2} D. Bernard and B. Doyon, Ann. Henri Poincar\'e 16, 113 (2015).
\bibitem{3} D.~Bernard and B.~Doyon,
J. Stat. Mech. \textbf{1606}, no.6, 064005 (2016)

\bibitem{4}  S. Hollands, R. Longo, Commun. Math. Phys. 2018, 357, 43.
\bibitem{pilo}
A.~Liguori, M.~Mintchev and L.~Pilo,
Nucl. Phys. B \textbf{569}, 577-605 (2000) 
\bibitem{mint0}
M.~Mintchev and P.~Sorba,
J. Phys. A \textbf{46}, 095006 (2013)
doi:10.1088/1751-8113/46/9/095006
[arXiv:1210.5409 [math-ph]].
\bibitem{mint}
M.~Mintchev and P.~Sorba,
Annalen Phys. \textbf{532}, no.10, 2000276 (2020)



\bibitem{thirring} W. Thirring, Ann. Phys. 3, 91 (1958).
\bibitem {wigcar}  
A.S. Wightman, 1964 Carg\`ese Summer School Lectures, 171-192. 
M. Levy, Ed., Gordon and Breach, New York (1967).


\bibitem{abdalla} E. Abdalla, M.C.B.  Abdalla  and  K.D. Rothe,  {\em Non Perturbative Methods in 2-Dimensional Quantum Field Theory}.   World Scientific, Singapore  (2001).
\bibitem{wigstr} 
F.~Strocchi and A.~S.~Wightman,
J. Math. Phys. \textbf{15}, 2198-2224 (1974)
[erratum: J. Math. Phys. \textbf{17}, 1930-1931 (1976)].
\bibitem{str} F.~Strocchi,
Phys. Rev. D \textbf{17}, 2010-2021 (1978)
\bibitem{johnson} K. Johnson, Nuovo Cimento 21, 773 (1961).
\bibitem{klaiber}
B.~Klaiber,
Lect. Theor. Phys. A \textbf{10}, 141-176, (1968).
\bibitem{strocchib}
G.~Morchio, D.~Pierotti and F.~Strocchi,
J. Math. Phys. \textbf{33} (1992), 777-790

\bibitem{mand} S. Mandelstam, 
Phys. Rev. D 11, 3026 (1975). 
\bibitem{cole} S. Coleman, 
Phys. Rev. {\bf D11}, 2088 (1975).




\bibitem{yoko}
H.~Yokota,
Prog. Theor. Phys. \textbf{77} (1987), 1450
[erratum: Prog. Theor. Phys. \textbf{81} (1989), 725]

\bibitem{sachs}
I.~Sachs, A.~Wipf and A.~Dettki,
Phys. Lett. B \textbf{317} (1993), 545-549
\bibitem{das}
A.~K.~Das and A.~J.~da Silva,
Phys. Rev. D \textbf{59} (1999), 105011
\bibitem{rothe}
R.~L.~P.~G.~Amaral, L.~V.~Belvedere and K.~D.~Rothe,
Annals Phys. \textbf{320}, 399-428 (2005)
\bibitem{kore} S.~E.~Korenblit and V.~V.~Semenov,
Russ. Phys. J. \textbf{55}, 1011-1021 (2013)




\bibitem{sbook}
F.~Strocchi,
``An introduction to non-perturbative foundations of quantum field theory,''
Int. Ser. Monogr. Phys. \textbf{158}, Oxford University Press (2013) 



\bibitem{strocchi}
G.~Morchio, D.~Pierotti and F.~Strocchi,
J. Math. Phys. \textbf{31} (1990), 1467


\bibitem{sw}
R. F. Streater and A. S. Wightman, "PCT, spin and statistics, and all that". Princeton: Princeton University Press, 2000 
\bibitem{gelfand}
I. M. Gel'fand and G. E. Shilov
"Generalized Functions", Volume 1.
Providence: AMS Chelsea Publishing, 1964 




\bibitem{jaffe}
A. Jaffe, Ann. Phys. \textbf{32} (1965), 127
\bibitem{pierotti}
D.~Pierotti,
Lett. Math. Phys. \textbf{15} (1988), 219



\bibitem{dashen} 
C.~G.~Callan, R.~F.~Dashen and D.~H.~Sharp,
Phys. Rev. \textbf{165}, 1883-1886 (1968)
doi:10.1103/PhysRev.165.1883
\bibitem{sugawara} H. Sugawara,
Phys. Rev. 170, 1659  (1968)
\bibitem{della}
G.~F.~Dell-Antonio, Y.~Frishman and D.~Zwanziger,
Phys. Rev. D \textbf{6}, 988-1007 (1972)
doi:10.1103/PhysRevD.6.988



\end{thebibliography}
\end{document}